\documentclass[preprint,onecolumn,floatfix,superscriptaddress,amsmath,amsfonts,amssymb,aps]{revtex4}

\usepackage{graphicx}
\usepackage{dcolumn}
\usepackage{bm}
\usepackage[normalem]{ulem}
\usepackage{color}
\usepackage{amssymb}
\usepackage{amstext}

\begin{document}


\title{Bi layer properties in the Bi--FeNi 
GMR-type structures \\ probed by spectroscopic ellipsometry}

\author{Natalia Kovaleva}
\email{kovalevann@lebedev.ru}
\affiliation{Lebedev Physical Institute, Russian Academy of Sciences\\
\hspace{0.46cm} Leninsky prospect 53, Moscow 119991, Russia}
\author{Dagmar Chvostova}
\author{Ladislav Fekete}
\affiliation{Institute of Physics, Academy of Sciences of the Czech Republic\\
\hspace{0.46cm} Na Slovance 2, Prague 18221, Czech Republic}
\author{Alexandr Dejneka}
\email{dejneka@fzu.cz}
\affiliation{Institute of Physics, Academy of Sciences of the Czech Republic\\
\hspace{0.46cm} Na Slovance 2, Prague 18221, Czech Republic}

\date{\today}

\begin{abstract}
Bismuth (Bi) having a large atomic number 
is characterized by the strong spin-orbit coupling (SOC) and is 
a parent compound of many 3D topological insulators (TIs). 
The ultrathin Bi films are supposed to be 2D TIs possessing 
the nontrivial topology, which opens the possibility of 
developing new efficient technologies in the field of spintronics. 
Here we aimed at studying the dielectric function properties of 
ultrathin Bi/FeNi periodic structures using spectroscopic ellipsometry. 
The [Bi($d$)--FeNi(1.8 nm)]$_{\rm N}$ GMR-type structures 
were grown by rf sputtering deposition on Sitall-glass (TiO$_2$) substrates. The ellipsometric angles $\Psi(\omega)$ and $\Delta(\omega)$ were measured for the grown series ($d$=0.6,\,1.4,\,2.0,\,2.5\,nm, N\,=\,16) 
of the multilayered film samples at room temperature for 
four angles of incidence of 60$^\circ$, 65$^\circ$, 70$^\circ$, and 
75$^\circ$ in a wide photon energy range of 0.5-6.5\,eV. 
The measured ellipsometric angles, $\Psi(\omega)$ and $\Delta(\omega)$, were simulated in the framework of the corresponding multilayer model. 
The complex (pseudo)dielectric function spectra of the Bi layer 
were extracted. The GMR effects relevant for the studied Bi--FeNi MLF systems are estimated from the optical conductivity zero-limit (optical GMR effect). The obtained results demonstrate that the Bi layer possesses the 
surface metallic conductivity induced by the SOC effects, which is 
strongly enhanced on vanishing the semimetallic-like phase contribution 
on decreasing the layer thickness, indicating its nontrivial 
2D topology properties.
\end{abstract}

\pacs{Valid PACS appear here}
\maketitle
\section{Introduction}
The relativistic effect of spin--orbit (SOC) coupling is involved in 
the so-called Rashba effect~\cite{Rashba}. This phenomenon arises 
from the apparent loss of crystalline inversion symmetry near the 
surface or heterojunction leading to the lifting of the spin degeneracy 
and generating spin-polarized surface metallic states. In~this respect, 
3D (2D) topological insulators (TIs) also exhibit spin-polarized 
surface metallic states due to SOC. However, contrary to the Rashba 
effect, the~surface metallic bands of a TI are determined by its 
bulk characteristics. The~TIs host metallic surface states in a bulk 
energy gap, which are topologically protected. The~surface 
(or interface) states of TIs can be topologically trivial or nontrivial. In~the latter case, for~example, electrons cannot be backscattered 
by impurities.
Bismuth (Bi) having a large atomic number is characterized by the strong SOC and is a parent compound of many 3D TIs, such as 
Bi$_{1-x}$Sb$_x$ or Bi$_2$Se$_3$, even although 3D bulk Bi itself 
is topologically trivial. The specific feature of the electronic 
band structure of bulk Bi having $R{\bar3}m$ rhombohedral symmetry 
\cite{Golin,Gonze,Liu1} is its inverted band gaps at both the 
$\Gamma$ and $M$ points of the Brillouin zone due to the strong 
SOC. The uniqueness of Bi films associated with the surface metallic 
states \cite{Hofmann,Yokota} and the semiconductor-to-metal transition 
\cite{Hoffman,Koroteev} are well documented in the literature.      

Theoretical analyses predict a 1-bilayer (BL) Bi(111) film to be 
a 2D TI \cite{Wada,Murakami}. If there is no or weak inter-BL 
coupling, a stack of the odd-even 1-BL films will exhibit nontrivial 
to trivial oscillations of topology (where the topological number 
$\nu$ \cite{Kane} is equal to 1 or 0, respectively). However, for 
the nontrivial topology in a stack of the 1-BL films, the intermediate 
inter-BL coupling strength, which is, for example, higher than the 
van der Waals strengths, is a mandatory condition. The direct ($\Gamma$ 
point) and indirect band gap values were calculated by Liu {\it et al.} 
as a function of the Bi film thickness \cite{Liu2}. It was established 
that below 4BLs the film is a semiconductor with the direct gap open 
at the $\Gamma$ point and the positive indirect band gap leading to 
nontrivial topology peculiar for an intrinsic 2D TI. Above 4BLs 
the indirect band gap becomes negative resulting in a semiconductor-
semimetal transition due to overlapping of two bands at the Fermi 
level around the $\Gamma$ and $M$ points. This suggests that the Bi 
films from 5 to 8 BLs represent a 2D TI situated between two trivial 
metallic surfaces \cite{Liu2}. 

A comprehensive study of the associated SOC effects in ultrathin Bi 
layers opens the possibility of developing new efficient technologies 
in the field of spintronics. For this purpose, here we aimed at 
studying the dielectric function properties of ultrathin periodic 
structures Bi/Ni$_{79}$Fe$_{21}$, prepared by rf sputter deposition, 
which is one of the most common technologies used to grow coatings and 
multilayered films (MLFs) exhibiting a giant magnetoresistance (GMR) 
effect for various existing and modern nanotechnological applications. 
Earlier, we have demonstrated that electronic band structure and surface electronic properties of ultrathin Bi layers in real GMR-type (Bi--FeNi)$_{\rm N}$ MLF structures incorporating nanoisland FeNi layers can be successfully studied by spectroscopic ellipsometry (SE) \cite{Kovaleva_APL_2021}. Here, by applying the elaborated SE approach, we investigate (Bi--FeNi) MLFs, where the thickness of the FeNi layer was 1.8\,nm, corresponding to the FeNi layer structural percolation threshold \cite{Sherstnev,Boltaev}, and the Bi spacer layer was 0.6, 1.4, 2.0, and 2.5\,nm thick, incorporating about 2, 4, 6, and 8 Bi$\left( 012 \right)$-type planes, respectively. We found that the Bi spacer layers have the metallic surface conductivity, which demonstrates strongly enhanced metallicity properties on vanishing the Bi semimetallic-like phase contribution on decreasing the layer thickness, which can be constructive in finding new nontrivial 2D topology properties of the (Bi--FeNi) GMR-type structures for their different nanotechnological applications.

\section{Materials and Methods}
The (Bi--FeNi)$_{\rm N}$ MLFs were prepared in a sputter deposition 
system by cathode sputtering from 99.95\% pure Bi and Fe$_{21}$Ni$_{79}$ targets in an alternative way. The base pressure in a sputter deposition chamber was 2$\times$10$^{-6}$ Torr. The multilayers were deposited 
at approximately 80\,$^{\circ}$C in an argon atmosphere of 6$\times$
10$^{-4}$ Torr on to insulating glassy Sitall (TiO$_2$) substrates. 
We utilized the substrates having typical dimensions 15\,$\times$\,5\,$\times$\,0.6\,mm$^3$. The nominal thicknesses of the FeNi and Bi layers were controlled by the layer deposition times in accordance with the material deposition rates. A series consisting of four MLF samples was prepared. In the series of the grown (Bi--FeNi)$_{\rm N}$ samples, the nominal thickness of the FeNi layer was 1.8\,nm and the Bi layer thickness was of 0.6, 1.4, 2.0, and 2.5\,nm, the number N of the periodically repeated Bi/FeNi layers was 16. The thickness of the FeNi layer was chosen to be 1.8\,nm matching the structural percolation threshold \cite{Sherstnev,Boltaev}. The Bi layer thicknesses were chosen in such a way that the conditions for ferromagnetic (FM) or antiFM 
coupling in the GMR-type structures would be optimized. 
To prevent degradation, the deposited 
(Bi--FeNi)$_{16}$/Sitall samples were covered with the 2.1\,nm-thick Al$_2$O$_3$ layer. 
\begin{figure}
\centering
\includegraphics[width=13.0cm]{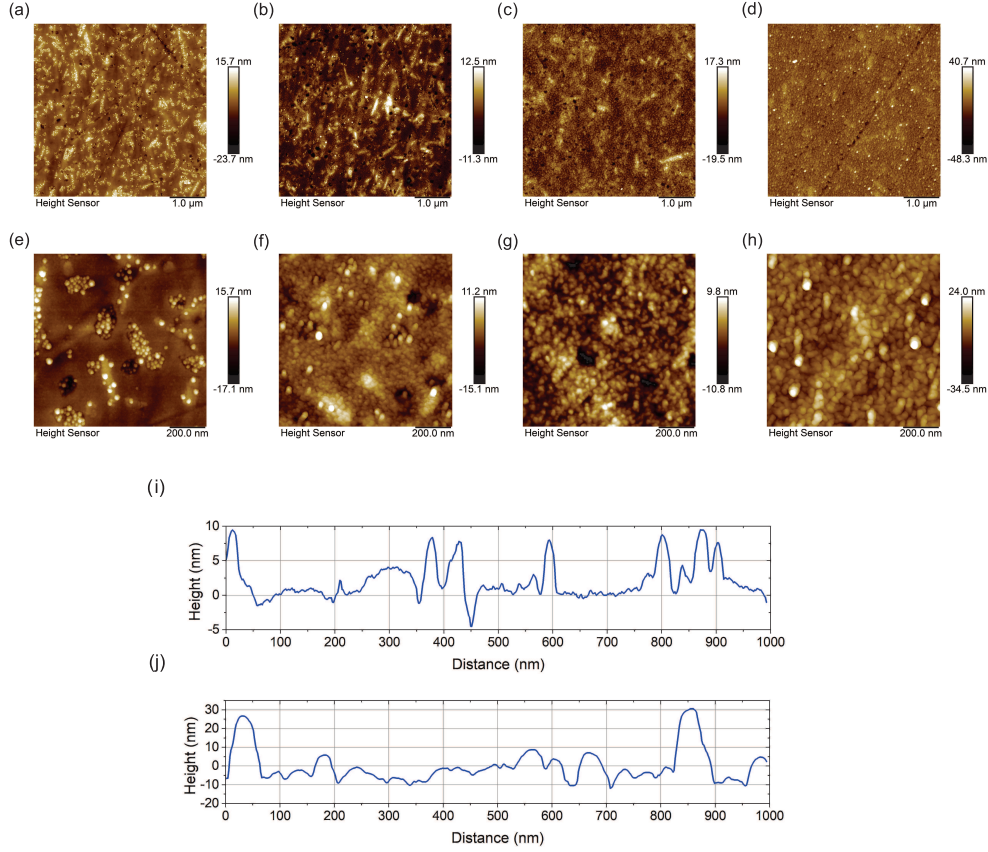}
\caption{AFM images {\bf (a--d)} 5\,$\times$\,5\,$\mu$m$^2$ and 
{\bf (e--h)} 1\,$\times$\,1\,$\mu$m$^2$ of the Al$_2$O$_3$/(Bi--FeNi)$_{16}$/Sitall MLF samples, where the nominal Al$_2$O$_3$ and FeNi layer thicknesses are 2.1 and 1.8\,nm and the nominal Bi layer thicknesses are 0.6, 1.4, 2.0, and 2.5\,nm, respectively. The estimated surface RMS roughness values are in {\bf (a-d)} 3.6, 3.0, 3.1, and 5.2\,nm and in {\bf (e-h)} 3.2, 2.6, 2.7, and 5.3\,nm, respectively. {\bf (i,j)} The typical height profiles for the MLF samples with the nominal Bi layer thicknesses of 0.6 and 2.5\,nm, respectively.}
\label{AFMimage}
\end{figure}
The related [Bi--FeNi(0.8,1.2\,nm)]$_{\rm N}$ samples prepared by 
rf sputtering deposition onto the Sitall substrates under similar 
conditions were investigated by X-ray diffraction (XRD) as well as 
by X-ray reflectivity (XRR) experimental techniques in our previous 
study (see Supplementary online information for the article 
\cite{Kovaleva_APL_2021}). 
The XRR spectra proved a good periodicity and consistency with the 
corresponding nominal thicknesses of the FeNi and Bi slices in the 
Bi/FeNi MLF structures, as well as relatively low interface 
roughness between the constituent layers. The XRD characterization 
suggests (012)-type Bi plane orientation, where the interlayer 
distance is 3.28\,\AA. It follows from this that in the studied 
MLF structures the Bi layers with a thickness corresponding to 0.6, 1.4, 2.0, and 2.5\,nm incorporate two, four, six, and eight Bi(012)-type 
planes, respectively. 

In the present study, the surface morphology of the Bi--FeNi(1.8\,nm) MLF samples, prepared by rf sputtering deposition on the Sitall (TiO$_2$) substrates, was studied at room temperature using an ambient AFM 
(Bruker, Dimension Icon) in the Peak Force Tapping mode with ScanAsyst Air tips (Bruker, k=0.4 N/m, nominal tip radius 2 nm).
The SE measurements for the investigated Al$_2$O$_3$/(Bi--FeNi)$_{16}$/Sitall samples were performed at room temperature in a wide photon energy 
range of 0.5\,--\,6.5\,eV using a J.A. Woollam VUV-VASE ellipsometer 
(see the scheme illustrating the SE study of the (Bi--FeNi)$_{\rm N}$ MLFs in Fig.\,1(a) of Ref.\,\cite{Kovaleva_APL_2021}). The measured ellipsometry spectra are represented by real values of the angles $\Psi(\omega)$ and $\Delta(\omega)$, which are defined through the complex Fresnel reflection coefficients for light-polarized parallel $r_p$ and perpendicular $r_s$ to the plane of incidence, ${\rm tan}\,\Psi\,e^{i\Delta}=\frac{r_p}{r_s}$. The ellipsometric angles, $\Psi(\omega)$ and $\Delta(\omega)$, measured for the Bi--FeNi MLF samples were simulated using the multilayer model simulation available in the 
J.A. Woollam VASE software \cite{VASE}. From the multilayer model simulations, the (pseudo)dielectric function spectra of the ultrathin 0.6,\,1.4,\,2.0, and 2.5\,nm Bi layers and 1.8\,nm FeNi layer inside the Bi--FeNi MLF structures were extracted. The corresponding calculated optical conductivity spectra were analyzed.     

\section{Results}
\subsection{Atomic force microscopy study}
The retrieved 5$\times$5\,$\mu$m$^2$ and 1$\times$1\,$\mu$m$^2$ AFM 
images of the Al$_2$O$_3$(2.1\,nm)/[Bi(0.6, 1.4, 2.0, 2.5\,nm)--FeNi(1.8\,nm)]$_{\rm N}$/Sitall multilayered films (where the given layer thicknesses correspond to their nominal values), presented in Figure\,\ref{AFMimage}a--h show 
discernable contrast because of the available surface hight deviations. The surface roughness of the Sitall glass (TiO$_2$) substrates was 
investigated by us by AFM in our earlier publication \cite{Stupakov}. The height profile of the Sitall substrates (see Fig.\,2a of Ref.\,\cite{Stupakov}) demonstrated the height deviation within the range 1-3 nm peculiar to the relatively large 0.3-1 $\mu$m lateral scale, which characterizes 
the Sitall substrate surface roughness. 
From the AFM measurements on the areas 5$\times$5 $\mu$m$^2$ and 1$\times$1 $\mu$m$^2$ the root-mean square (RMS) surface roughness values were evaluated, which are presented in the caption to Figure\,\ref{AFMimage}. The corresponding RMS roughness values are notably higher for the 
Al$_2$O$_3$(2.1\,nm)/[Bi(2.5\,nm)--FeNi(1.8\,nm)]$_{16}$/Sitall MLF sample. The smaller-scale (1\,$\times$\,1\,$\mu$m$^2$) images clearly recognize a fine grainy-like structure of the surface morphology, which seems to be characteristic for all studied  film samples (see Figure\,\ref{AFMimage}e--h). 
The typical grain size, being of about 50\,nm, is notably larger for the FeNi(1.8\,nm) -- Bi MLF sample incorporating the 2.5 nm-thick Bi layers, and, following the estimated RMS roughness values, the average grain size decreases to about 20\,nm with decreasing the Bi layer thickness to 1.4\,nm. As one can see from the typical height profiles presented in Figure\,\ref{AFMimage}i,j, with decreasing the Bi layer thickness from 2.5 to about 0.6\,nm, the surface morphology becomes highly irregular due to the formation of conglomerates of nanoislands separated by rather flat (relatively small roughness) areas of about 20\,nm. 

\subsection{\mbox{Spectroscopic ellipsometry study of the ultrathin 
Bi--FeNi multilayer film samples}}
\begin{figure}
\centering
\includegraphics[width=8.0cm]{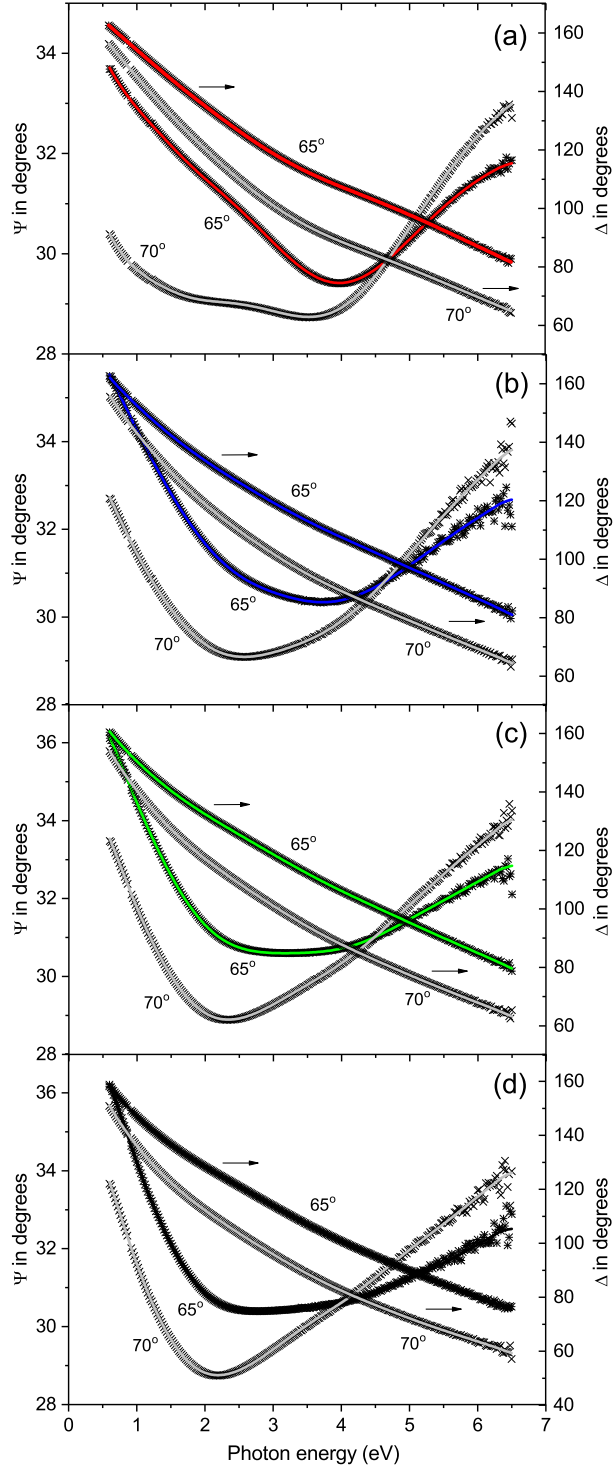}
\caption{{\bf (a-d)} Ellipsometric angles, $\Psi(\omega)$ and $\Delta(\omega)$ (symbols), measured at the angles of incidence of 65$^\circ$ and 
70$^\circ$ for the Al$_2$O$_3$/[Bi($d$)--NiFe(1.8\,nm)]$_{16}$/Sitall multilayered films where the Bi spacer layer thicknesses $d$\,=\,0.6, 1.4, 2.0, 
and 2.5\,nm, respectively. The solid red, blue, green, and black 
curves show the corresponding simulation results for the 
angle 65$^\circ$ by the dielectric function model using Equation\,\ref{DispAna}.}
\label{PsiDelta}
\end{figure}
\begin{figure}
\centering
\includegraphics[width=7.5 cm]{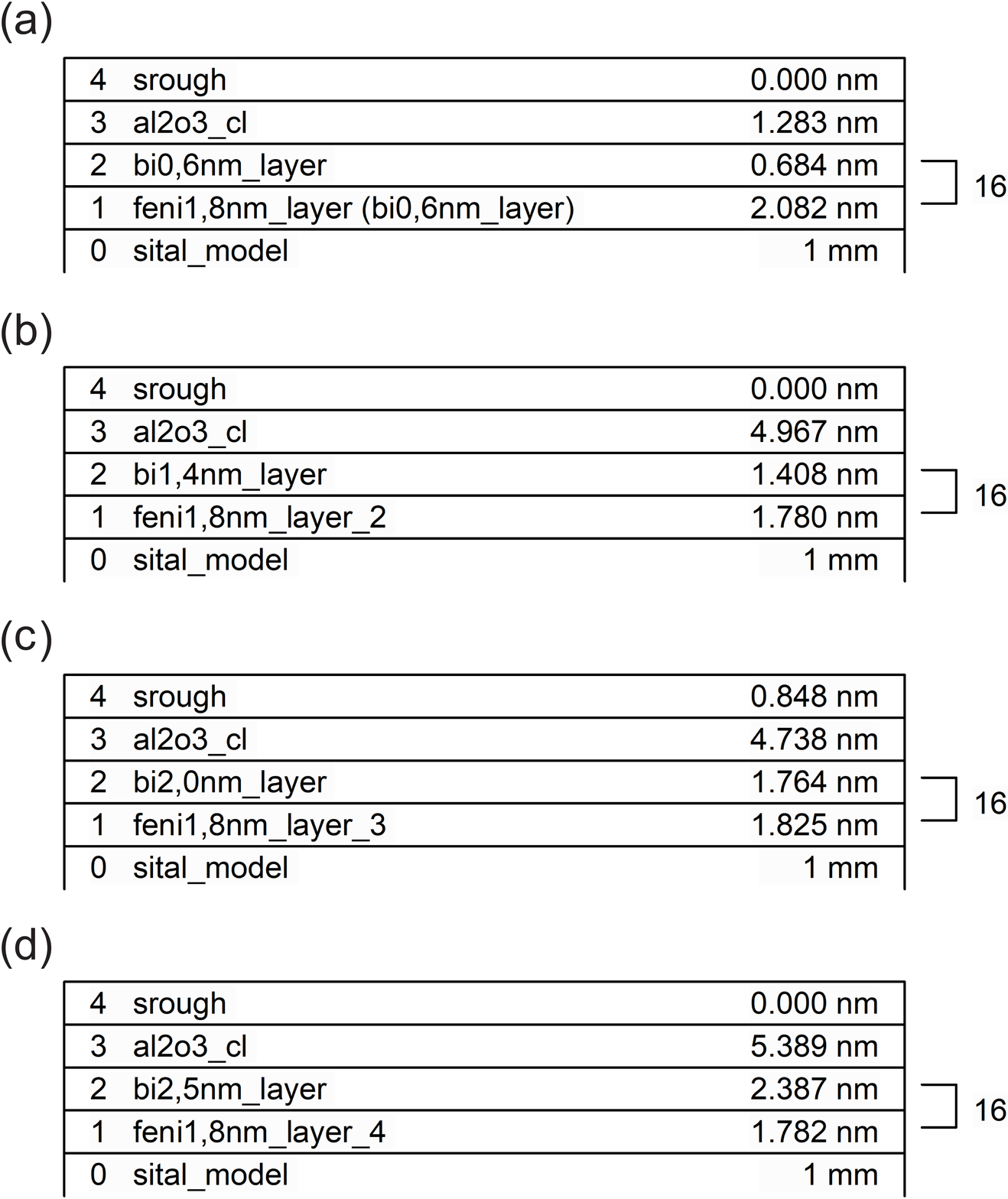}
\caption{ The multilayer model applied for the simulation 
of the Al$_2$O$_3$/[Bi(0.6, 1.4, 2.0, and 2.5\,nm)--FeNi(1.8\,nm)]$_{16}$/Sitall samples. The Bi and FeNi thicknesses estimated from the model 
simulations in ({\bf a}) 0.684$\pm$0.037\,nm and 2.082$\pm$0.116\,nm, 
({\bf b}) 1.408$\pm$0.574\,nm and 1.780$\pm$0.65\,nm, 
({\bf c}) 1.764$\pm$0.194\,nm and 1.825$\pm$0.358\,nm, 
and ({\bf d}) 2.387$\pm$0.128\,nm and 1.782$\pm$0.171\,nm. 
Note good agreement between the thicknesses 
of the FeNi and Bi layers estimated from the model simulations and 
their respective nominal thickness values. The roughness and Al$_2$O$_3$ thicknesses estimated from the model 
simulations in ({\bf a}) 0.00$\pm$3.85\,nm and 1.283$\pm$2.37\,nm, 
({\bf b}) 0.000$\pm$4.97\,nm and 4.967$\pm$2.17\,nm, 
({\bf c}) 0.848$\pm$5.86\,nm and 4.738$\pm$2.92\,nm, 
and ({\bf d}) 0.000$\pm$2.95\,nm and 5.389$\pm$1.23\,nm.}
\label{Model}
\end{figure}
\begin{figure}
\centering
\includegraphics[width=10.0cm]{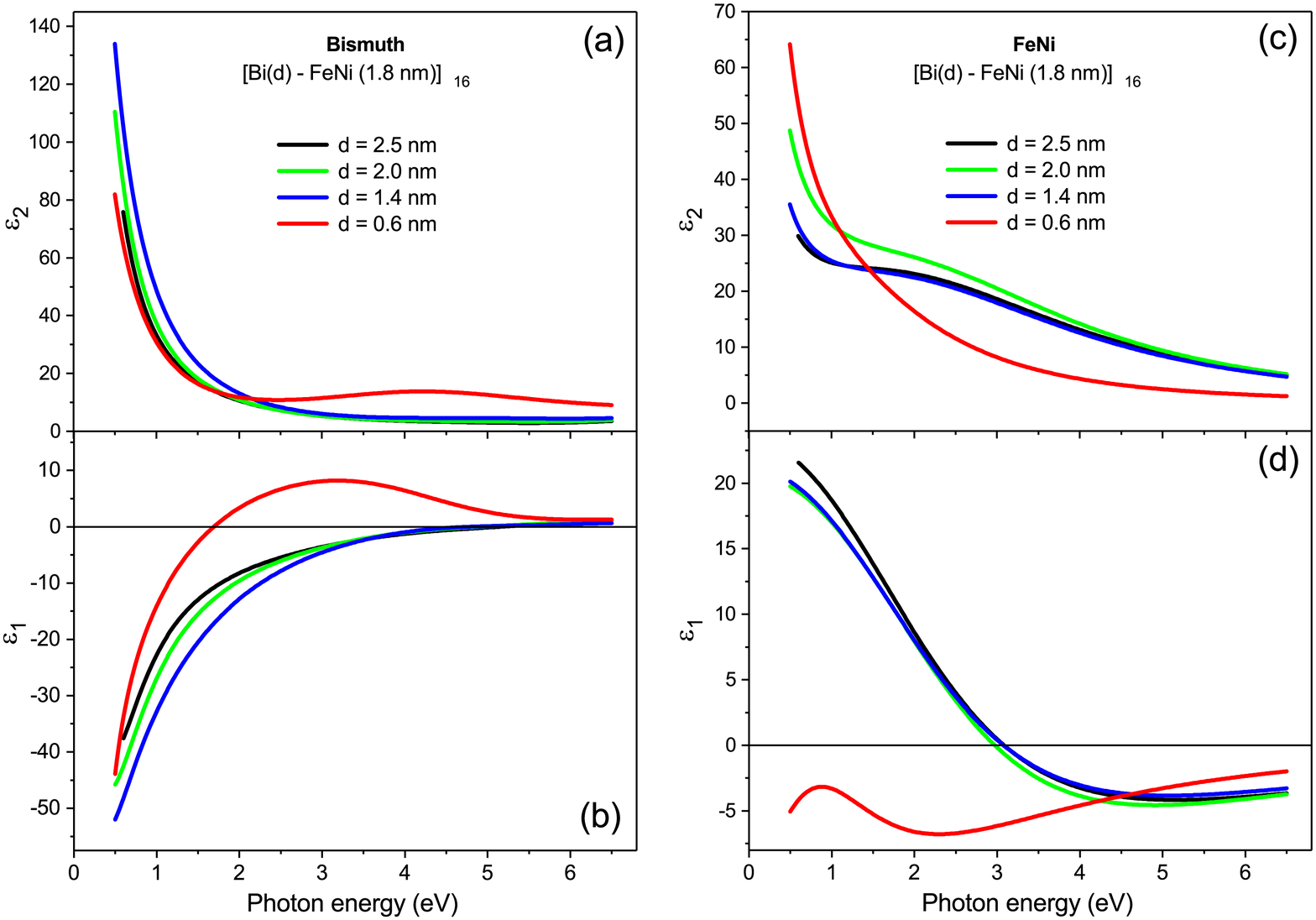}
\caption{The complex (pseudo)dielectric function spectra, $\varepsilon_2(\omega)$ and $\varepsilon_1(\omega)$, of the ({\bf a,b}) Bi layers and ({\bf c,d}) FeNi layers in the [Bi($d$)--FeNi(1.8\,nm)]$_{16}$ structures shown for 
the Bi layer nominal thickness values $d=$ 0.6, 1.4, 2.0, and 2.5\,nm 
by solid red, blue, green, and black curves, respectively.}
\label{e1e2}
\end{figure}

The ellipsometric angles $\Psi(\omega)$ and $\Delta(\omega)$ were 
measured for the prepared Al$_2$O$_3$/(Bi--FeNi)$_{16}$/Sitall MLF 
samples at the angles of incidence of 60$^\circ$, 65$^\circ$, 70$^\circ$, and 75$^\circ$. Figure\,\ref{PsiDelta} demonstrates the ellipsometric angles $\Psi(\omega)$ and $\Delta(\omega)$ recorded at 65$^\circ$ and 
70$^\circ$. To model the contributions from free charge carriers and 
interband optical transitions, the complex dielectric function 
$\tilde \varepsilon(\omega)=\varepsilon_1(\omega)+{\rm i} \varepsilon_2(\omega)$ of the Bi and FeNi layers was interpreted in terms of the Drude  
and Lorentz parts, respectively, 
\begin{eqnarray}
\tilde\varepsilon(E\equiv \hbar\omega)=\epsilon_{\infty}-\frac{A_D}{E^2+{\rm i}E\gamma_D}+\sum_j\frac{A_j \gamma_jE_j}{E_j^2-E^2-{\rm i}E\gamma_j}, 
\label{DispAna}
\end{eqnarray}
where $\varepsilon_{\infty}$ is the high-frequency dielectric constant, 
which takes into account the contribution from the higher-energy 
interband transitions. The fitted Drude parameters were $A_D$ and free charge carrier's scattering rate $\gamma_D$. The fitted parameters of Lorentz bands were $E_j$, $\gamma_j$, and $A_j$ of the band maximum energy, the full width at half maximum, and the $\varepsilon_2$ band height, respectively. 
The obtained ellipsometric angles $\Psi(\omega)$ and $\Delta(\omega)$ 
measured at different angles of incidence of 60$^\circ$, 65$^\circ$, 
70$^\circ$, and 75$^\circ$ were fitted for each sample simultaneously 
using the J.A. Woollam VASE software \cite{VASE} in the framework 
of the designed multilayer model. The multilayer model for the studied 
Al$_2$O$_3$/(Bi--FeNi)/Sitall multiayers was constructed as it is schematically presented in Figure\,\ref{Model}, exactly so as the layers were 
deposited. In addition, we attempted to take into account the roughness properties of the surface by using the conventional approach of effective medium approximation (EMA) based on the (50\% Al$_2$O$_3$--50\% vacuum) Bruggeman model. The dispersion model for the Bi layers included three or four Lorentz terms as well as the Drude part. The dispersion model for the 1.8\,nm 
permalloy layers incorporated in the studied MLF structures included the Drude term responsible for the free charge carrier contribution and one Lorentz oscillator to account for the most pronounced interband optical transition.
In addition, the dielectric function spectra of the bare Sitall substrate derived from our earlier SE studies \cite{Kovaleva_APL_2015,Kovaleva_metals} were introduced to the elaborated multilayer model. The dielectric 
response of the Al$_2$O$_3$ capping layer was represented by the tabular complex dielectric function spectra \cite{Palik}. The thicknesses of 
the Bi and FeNi layers, as well as of the surface layers, were fitted. 
The unknown parameters were allowed to vary until the minimum of the mean squared error (MSE) is reached. The best simulation result for the studied [Bi(0.6,\,1.4,\,2.0,\,2.5\,nm)--FeNi(1.8\,nm)]$_{16}$ MLF samples corresponded to the lowest obtained MSE values of 0.3843, 0.297, 0.2934, and 0.4508, respectively.
The good quality of the fit allowed us to estimate the actual Bi and 
FeNi layer thicknesses in the MLFs under study. The quality of the fit 
is demonstrated by Figure\,\ref{PsiDelta}, where we plotted the measured ellipsometric angles along with the simulation results. The Drude 
and Lorentz parameters resulting from the simulation of the 
Al$_2$O$_3$/[Bi($d$)--FeNi(1.8\,nm)]$_{16}$/Sitall 
MLF samples are given in Tables\,\ref{Table1} and \ref{Table2}, 
and the resulting $\varepsilon_1(\omega)$ and $\varepsilon_2(\omega)$ parts of the Bi and FeNi (pseudo)dielectric function spectra are presented 
in Figure\,\ref{e1e2}. 

From Figure\,\ref{e1e2}a,b one can see that the complex (pseudo)dielectric functions of the 0.6, 1.4, 2.0, and 2.5 nm thick Bi spacers inside 
the investigated Bi--FeNi MLFs demonstrate metallic character. Moreover, the $\varepsilon_1(\omega)$ function progressively decreases 
while the Bi thickness decreases from 2.5--2.0 to 1.4\,nm and  
the $\varepsilon_2(\omega)$ increases at low photon energies, respectively. According to our simulation results, we expect that the best metallicity properties are demonstrated by the Bi layer in the 
[Bi(1.4\,nm)--NiFe(1.8\,nm)]$_{16}$ structure. At the same time, the complex
(pseudo)dielectric functions of the thinnest 0.6\,nm thick Bi 
layer look somewhat different. Here, in addition to the low-energy 
metallic Drude response identified by the characteristic behavior 
of the $\varepsilon_1(\omega)$ and $\varepsilon_2(\omega)$, the Lorentz 
band around 4--5\,eV makes an essential contribution to the dielectric function response (the corresponding Drude ($A_D$ and $\gamma_D$) and 
Lorentz ($A_j$, $E_j$, and $\gamma_j$) parameters are listed in Table\,\ref{Table1}). Next, being similar, the dielectric functions of the 1.8\,nm thick permalloy layers in the [FeNi--Bi(1.4,\,2.0,\,2.5\,nm)] MLFs are dominated 
by the $\varepsilon_2(\omega)$ resonance and $\varepsilon_1(\omega)$ 
antiresonance features, indicating the predominant contribution from 
the Lorentz oscillator peaking at around 3\,eV (see Figure\,\ref{e1e2}c,d). An upturn evident in the $\varepsilon_2(\omega)$ at low photon energies 
indicates an additional Drude contribution, which is relatively less 
pronounced. Following our simulation results, we expect the advanced 
metallicity properties of the FeNi layer in the 
[Bi(0.6\,nm)--NiFe(1.8\,nm)]$_{16}$ structure (see the corresponding Drude ($A_D$ and $\gamma_D$) and Lorentz ($A_j$, $E_j$, and $\gamma_j$) 
parameters listed in Table\,\ref{Table2}). 
\begin{table}
\centering
\caption{Drude-Lorentz parameters for the Bi spacer layer 
in the [Bi(0.6,\,1.4,\,2.0,\,2.5\,nm)--NiFe(1.8\,nm)]$_{16}$ multilayered films obtained from the model simulations of the dielectric functions by using Equation\,\ref{DispAna}. 
The values of $E_j$, $\gamma_j$, 
and $\gamma_D$ are given in eV, and optical conductivity limit 
$\sigma_{1(\omega\rightarrow0)}$ in $\Omega^{-1}\cdot$cm$^{-1}$.}
\begin{tabular}[c]{@{}l@{~~}l@{~~}l@{~~}l@{~~}l@{~~}l@{}}
\hline
&  {\bf Parameters}&{{\bf 0.6\ nm}}~&{{\bf 1.4\ nm}}~&{{\bf 2.0\ nm}}~&{{\bf 2.5\ nm}}\\ \hline
Drude & \quad$A_D$ & 46.(9)$\pm$4 & 66.(7)$\pm$4 &24.(5)$\pm$4 &25.(1)$\pm$2\\
& \quad$\gamma_D$  & 1.2(5)$\pm$0.09 &1.51(0)$\pm$0.06 &2.7(2)$\pm$0.4 & 3.1(3)$\pm$0.2\\ & \quad$\sigma_{1(\omega\rightarrow 0)}$ & 6300$\pm$540&8970$\pm$540& 3290$\pm$540& 3370$\pm$270\\ \hline
Lorentz &\quad$E_1$  & -- &0.45(8)$\pm$0.05  & 0.35(9)$\pm$0.01  &0.38(6)$\pm$0.004\\
oscillator & \quad$A_1$ & -- & 15.(0)$\pm$6  & 96.(0)$\pm10$ &70.(8)$\pm$2\\
& \quad$\gamma_1$ & -- & 0.52(6)$\pm$0.09 & 0.79(1)$\pm$0.02  &0.67(6)\\ \hline
Lorentz& \quad$E_2$ & 4.67   &5.31(5)$\pm$0.03&5.08(7)$\pm$0.04 & 4.77(5)$\pm$0.04 \\
oscillator& \quad$A_2$& 10.2(7)$\pm$0.6 &2.53(2)$\pm$0.05& 1.2(5)$\pm$0.1 & 0.67(6)$\pm$0.08\\ 
& \quad$\gamma_2$ & 4.2(1)$\pm$0.07 &3.99(3)$\pm$0.07&3.4(7)$\pm$0.2 & 2.5(5)$\pm$0.2\\ \hline
Lorentz& \quad$E_3$ & 11.1    &7.8 &7.7  & 7.7\\
oscillator& \quad$A_3$ & 7.2       &4.1& 4.1 & 4.1\\
& \quad$\gamma_3$ & 8.9  &2.8& 2.8 & 2.8\\ \hline
\end{tabular}
\label{Table1}       
\end{table}
\begin{table}
\centering
\caption{Drude-Lorentz parameters for the 1.8\,nm thick NiFe layer 
in the [Bi(0.6,\,1.4,\,2.0,\,2.5\,nm)--NiFe]$_{16}$ multilayered films obtained from the simulations of the model dielectric function described by Equation\,\ref{DispAna}. The values of $E_1$, $\gamma_1$, and $\gamma_D$ are given in eV, and optical conductivity limit 
$\sigma_{1(\omega\rightarrow0)}$ in $\Omega^{-1}\cdot$cm$^{-1}$.}
\begin{tabular}[c]{@{}l@{~~}l@{~~}l@{~~}l@{~~}l@{~~}l@{}}
\hline
&  {\bf Parameters}&{{\bf 0.6\ nm}}~&{{\bf 1.4\ nm}}~&{{\bf 2.0\ nm}}~&{{\bf 2.5\ nm}}\\ \hline
Drude & \quad$A_D$ & 33.(8)$\pm$2 & 15.(0)$\pm$1 &21.(7)$\pm$2 &13.(1)$\pm$2\\
& \quad$\gamma_D$  & 0.876(5)$\pm$0.04 &2.8(2)$\pm$0.3 &3.4(2)$\pm$0.4 & 3.1(3)$\pm$0.2\\ & \quad$\sigma_{1(\omega\rightarrow 0)}$ & 4540$\pm$270&2020$\pm$130& 2920$\pm$270& 1760$\pm$270\\ \hline
Lorentz &\quad$E_1$  & 1.87 &3.32  & 3.32  &3.32\\
oscillator & \quad$A_1$ & 14.76 & 14.28  & 15.23 &14.74\\
& \quad$\gamma_1$ & 3.62& 5.88 & 5.65  &5.95\\ \hline
\end{tabular}
\label{Table2}       
\end{table}

\begin{figure}
\centering
\includegraphics[width=12cm]{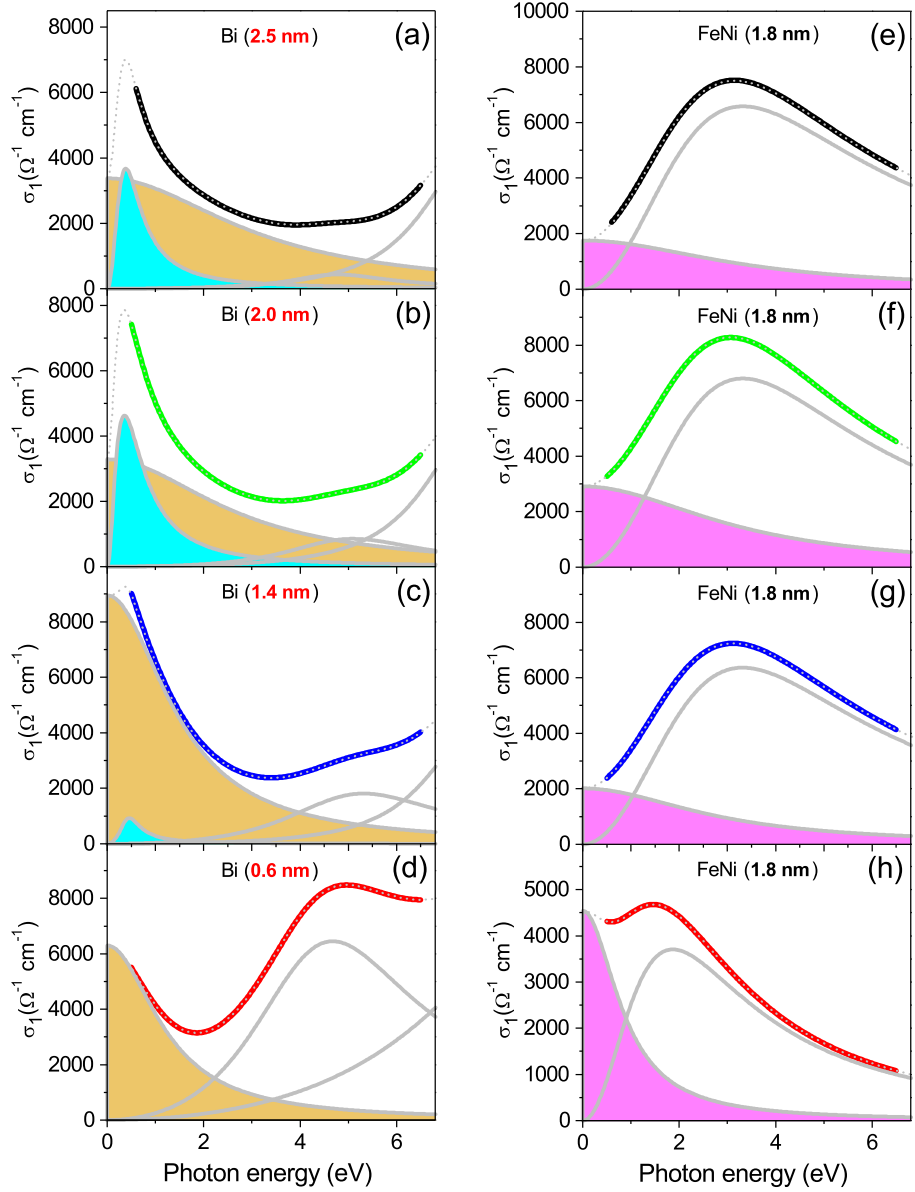}
\caption{The intralayer optical conductivity, 
$\sigma_1(\omega)=\varepsilon_2(\omega)\omega[{\rm cm^{-1}]/60}$, 
for the ({\bf a-d}) Bi layers and ({\bf e-h}) FeNi layers in 
the [Bi($d$)--FeNi(1.8\,nm)]$_{16}$ structures shown for the Bi 
layer nominal thickness values $d=$ 2.5, 2.0, 1.4, and 0.6\,nm by 
solid curves {\bf (a,e)} black, {\bf (b,f)} green, {\bf (c,g)} blue, 
and {\bf (d,h)} red, respectively. The contributions from the Drude 
term and the Lorentz oscillator in {\bf(a-d)} are displayed by the 
yellow and cyan shaded area. 
In {\bf(e-h)} the Drude term for the FeNi layers is displayed by 
the magenta shaded area. Shown by the dotted curves are the summary 
of the Drude and Lorentz contributions.}
\label{Disp}
\end{figure}

Figure\,\ref{Disp}a--d presents the evolution of the Bi intralayer 
optical conductivity, $\sigma_1(\omega)=\varepsilon_2(\omega)\omega$(cm$^{-1}$)/60, upon decreasing the Bi spacer layer thickness in 
the [FeNi(1.8\,nm) -- Bi(2.5,\,2.0,\,1.4,\,0.6\,nm)]$_{16}$ structures, 
and Figure\,\ref{Disp}e--h shows the associated optical conductivity 
spectra of the 1.8\,nm FeNi permalloy layer. Here, 
the contributions from the Drude and Lorentz  oscillators following 
the multilayer model simulations using Equation\,\ref{DispAna} are 
evidently demonstrated. The optical conductivity spectra 
of the Bi and FeNi layers follow the main trends identified in 
their complex dielectric function spectra presented in Figure\,\ref{e1e2}.

\section{Discussion}
\begin{figure}
\centering
\includegraphics[width=7.0cm]{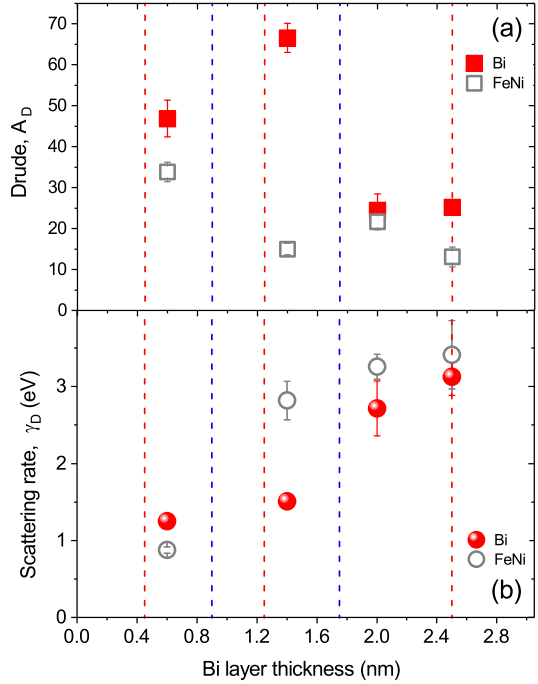}
\caption{({\bf a,b}) Parameters of the Drude term ($A_D$ and $\gamma_D$) for the Bi (filled symbols) and FeNi (empty symbols) layers in the 
[Bi(0.6,\,1.4,\,2.0,\,2.5\,nm)--FeNi(1.8\,nm)] MLF structures.}
\label{Drude}
\end{figure}

Initially, we would like to discuss GMR effects relevant for the 
studied MLF systems. Our simulations of the dielectric functions 
for the 1.8\,nm-thick NiFe layer inside the 
[Bi(0.6,1.4,2.0,2.5\,nm)--NiFe(1.8\,nm)] MLFs 
show the presence of the Drude term complemented with the pronounced 
Lorentz band located at around 2--3\,eV (see Table\,\ref{Table2}). 
From the corresponding optical conductivity spectra presented in 
Figure\,\ref{Disp}e--h one can notice that the associated Drude dc 
limit, $\sigma_{1\omega\rightarrow0}$, displays an oscillating character (in agreement with the results deduced for the corresponding Drude 
parameter $A_D$, see Table\,\ref{Table2} and Figure\,\ref{Drude}). We 
can expect that the Bi spacer thicknesses for which the FeNi layers 
are preferentially antiFM coupled in the studied 
MLFs are around 1.4 and 2.5\,nm implying that 
the [Bi(1.4,2.5\,nm)--NiFe(1.8\,nm)]$_{16}$ film structures 
will exhibit a drop in the resistance (being negative magnetoresistance) when exposed to an external magnetic field. It is well known from the 
literature that the first antiFM maximum exhibits negative magnetoresistance of about 20\%, while the second antiFM maximum decreases to about 10\%, 
and the presence of the third antiFM maximum cannot confidently be 
retrieved (see, for example, Ref.\,\cite{Huetten} and references 
therein). Using a simple model of a two-current series resistor 
\cite{Mathon}, the magnetoresistance $\frac{\Delta R}{R}$ can be 
estimated as
\begin{eqnarray}
\frac{\Delta R}{R}=100\%\frac{\left( \alpha -\beta \right)^2}{4\left( \alpha+\frac{d_{Bi}}{d_{FeNi}} \right)\left( \beta+\frac{d_{Bi}}{d_{FeNi}} \right)},
\label{GMR}
\end{eqnarray}
where $d_{Bi}$ and $d_{FeNi}$ are the thicknesses of Bi and FeNi layers, and $\alpha=\frac{^\downarrow\rho_{FeNi}}{\rho_{Bi}}$ and 
$\beta=\frac{^\uparrow\rho_{FeNi}}{\rho_{Bi}}$ are the ratios of the 
resistivity in the FeNi layer to that in the Bi layer in the spin down 
and spin up current channel, respectively. Exploiting values for 
$\rho=\sigma_{1\omega\rightarrow0}^{-1}$ estimated for the 1.4\,nm Bi and 1.8\,nm FeNi layers from the current model simulations (see Table\,\ref{Table1} and \ref{Table2}), namely, 
$\rho_{Bi}$=$\frac{1}{8970}$$\Omega\cdot$cm, 
$^\downarrow\rho_{FeNi}$=$\frac{1}{2020}$$\Omega\cdot$cm and 
$^\uparrow\rho_{FeNi}$=$\frac{1}{4540}$$\Omega\cdot$cm 
(the latter estimate is given by the FM coupling for the 0.6\,nm 
Bi spacer), we obtain $\alpha$=4.4 and $\beta$=2.0. Then, using Equation\,(\ref{GMR}) we have $\frac{\Delta R}{R}$=10\%. This means that the 1.4\,nm Bi spacer corresponds to the second antiFM maximum. Following the same approach for the 2.5\,nm Bi spacer, where $\rho_{Bi}$=$\frac{1}{3370}$$\Omega\cdot$cm, $^\downarrow\rho_{FeNi}$=$\frac{1}{1760}$$\Omega\cdot$cm and 
$^\uparrow\rho_{FeNi}$=$\frac{1}{2920}$$\Omega\cdot$cm 
(corresponding to the FM coupling for the 2.0\,nm Bi spacer), 
we obtain $\alpha$=1.9 and $\beta$=1.2. Using Equation\,(\ref{GMR}), 
we have 
$\frac{\Delta R}{R}$=1.4\%, which may correspond to the very weakly 
pronounced third antiFM maximum. From the analysis presented above, 
we may expect that the first antiFM maximum corresponding to the 
magnetoresistance of about 20\% occurs for the Bi spacer thickness 
of about 0.9\,nm, which is in agreement with the results presented in 
Ref.\,\cite{Huetten}.      

Further, in the XRD patterns of the investigated 
Al$_2$O$_3$/[Bi(1.4,2.0,2.5\,nm)--NiFe(1.8\,nm)]$_{16}$/Sitall film samples, the peak of the $R{\bar3}m$ crystalline Bi phase is identified at 2$\theta$\,$\approx$\,26.2$^\circ$ suggesting $\left( 012 \right)$ 
orientation of the Bi layers, which is characterized by the interlayer 
distance of 3.28\,\AA. Using STM and reflection high-energy electron 
diffraction (RHEED) techniques, it was shown that initial growth of 
Bi$\left( 012 \right)$-type films occurs in the form of islands with 
the height increment of about 6.6\,\AA, indicating even-number layer 
stability leading to the laterally flat morphology of the 
Bi$\left( 012 \right)$-type islands \cite{Nagao}. Consequently, we can 
expect that the 0.6,\,1.4,\,2.0,\,and 2.5\,nm Bi spacer layers in the investigated MLFs incorporate about 2, 4, 6, and 8 $\left( 012 \right)$-type Bi planes, respectively. 

The model simulations for the [Bi(2.5,\,2.0\,nm)--FeNi(1.8\,nm)]$_{16}$ film samples reveal that the low-energy dielectric function of the 
Bi intralayers has competing contributions from the Drude term and from the intense Lorentz band around 0.36--0.39\,eV with a $\varepsilon_2$ maximum height of 70--100 (see Table\,\ref{Table1}). The Drude and Lorentz contributions are more clearly pronounced in the corresponding optical 
conductivity spectra (see Figure\,\ref{Disp}a,b). The obtained Drude and Lorentz parameters are in excellent agreement with those deduced in our 
previous study \cite{Kovaleva_APL_2021} for the Bi spacer layer incorporated in the [Bi(2.5,\,2.0\,nm)--NiFe(1.2\,nm)]$_{16}$ structures under study. The pronounced Lorentz band found at low photon energies for Bi single 
crystals (rhombohedral symmetry, space group $R{\bar3}m$) \cite{Wang,Lenham} and bulk Bi layers \cite{Hunderi,Toudert} is characteristic of the 
semimetallic-like electronic band structure due to the contributions from the interband transitions near the $\Gamma$ point, $\Gamma^+_6$ -- $\Gamma^-_6$ and $\Gamma^+_{45}$ -- $\Gamma^-_6$ \cite{Golin}, and near the T point, 
T$^-_6$ -- T$_{45}^-$ \cite{Liu1}. The estimated values (see Table\,\ref{Table1}) of the Drude dc limit $\sigma_{1\omega\rightarrow 0}$ 
(2750--3830\,$\Omega^{-1}$$\cdot$cm$^{-1}$) as well as the free charge carrier's $\gamma_D$ (2.3--3.3\,eV) are consistent with those peculiar for the metallic surface states related to the Rashba SOC in Bi(111) films, 
$\sigma_{1\omega\rightarrow 0}$ = 2300\,$\Omega^{-1}$$\cdot$cm$^{-1}$ and $\gamma_D$\,=\,2.0\,eV) \cite{Yokota}. Meanwhile, the model simulation for the [Bi(1.4\,nm)--NiFe(1.8\,nm)]$_{16}$ structure indicates that for the 1.4\,nm Bi layer the Drude dc limit significantly increases to 
8970$\pm$540\,$\Omega^{-1}$$\cdot$cm$^{-1}$, while the $\gamma_D$ essentially decreases to 1.50$\pm$0.06\,eV. In this case, the Lorentz band is nearly suppressed. The associated found Drude parameters for the ultrathin Bi layer inside the [Bi(0.6\,nm)--NiFe(1.8\,nm)]$_{16}$ structure are slightly different, namely, $\sigma_{1\omega\rightarrow 0}$ = 6300$\pm$540\,$\Omega^{-1}$$\cdot$cm$^{-1}$ and $\gamma_D$ = 1.2$\pm$0.1\,eV, and the Lorentz band is not present clearly (see Figure\,\ref{Disp}c,d and Table\,\ref{Table1}).

Thus, we have discovered that, on the one hand, the optical conductivity spectra spectra of the 2.0 and 2.5\,nm thick Bi spacer layers in the (Bi--FeNi) MLFs incorporating 8 and 6 Bi$\left( 012 \right)$-type monolayers, respectively, have contributions from the pronounced low-energy Lorentz oscillator and from the free charge carrier Drude term (for details, see Figure\,\ref{Disp}a,b and Table\,\ref{Table1}). Here, the presence of the low-energy Lorentz band points on the Bi semimetallic phase contribution, and the parameters obtained for the Drude conductivity indicate that its origin can be associated with the surface metallic states \cite{Yokota}. Therefore, the 2.0 and 2.5\,nm Bi layers can be associated with the semimetallic Bi phase sandwiched between two metallic layers on the top and bottom surfaces. On the other hand, the contribution from the intrinsic Lorentz band is strongly suppressed for the 1.4 and 0.6\,nm layers, where the Drude conductivity displays notably improved metallicity properties, as one can see from the optical conductivity spectra shown in Figure\,\ref{Disp}c,d (for details, see Table\,\ref{Table1}). 

From the above discussion of the obtained results, we can conclude that the Bi layer consisting of 4 Bi(012)-type monolayers represents a kind of crossover regarding the contributions from the semimetallic Bi phase and/or surface metallic-like states. Here we noticed some similarity with the theory results presented for the ultrathin Bi(111) layers by Liu {\it et al.} \cite{Liu2}. There, it was established that below 4 Bi(111) BLs the film is a semiconductor with the direct gap open at the $\Gamma$ point and the positive indirect band gap, leading to nontrivial $Z_2$ topology ($\nu$=1) peculiar for an intrinsic 2D TI. Hovewer, above 4 Bi(111) BLs, the indirect band gap becomes negative resulting in a semiconductor-semimetal transition due to overlapping of two bands at the Fermi level around the $\Gamma$ and $M$ points. It is argued by Liu {\it et al.} \cite{Liu2} that the Bi layers consisting of 5 to 8 Bi(111) BLs represent a 2D TI suited between two ``trivial'' metallic surfaces \cite{Liu2}. This means that for the surface considered as an individual 2D system its $Z_2$ number is trivial ($\nu$=0). The surface bands have no contribution to the nontrivial $Z_2$ topology and, therefore, these trivial metallic surfaces are not robust and can easily be removed by surface defects or impurities. It was found by us \cite{Kovaleva_APL_2021} that the Bi layers in the [Bi(2.0,\,2.5\,nm)--NiFe(0.8\,nm)] multilayers, incorporating the nanoisland permalloy layer, exhibit bulk-like semimetallic properties of the electronic band structure, although the surface (Drude) metallic conductivity is absent there (see Fig.\,4(d) of Ref.\,\cite{Kovaleva_APL_2021}). Indeed, strong magnetic and spatial disorder induced by magnetic FeNi nanoislands, as well as long-range many-body interactions between magnetic moments of permalloy nanoislands \cite{Stupakov}, may lead to specific localization of free charge carriers \cite{Kovaleva_SciRep}. However, the surface conductivity (or interface) states for the 1.4 nm layer in the Bi--FeNi(1.8\,nm) multilayers may be topologically nontrivial and, in this case, the electrons cannot be backscattered by impurities. Here, the Drude dc limit is 
8970$\pm$540\,$\Omega\cdot$cm$^{-1}$ and the scattering rate 
$\gamma_D$=1.5$\pm$0.06\,eV. We found that the 0.6\,nm thick Bi layer 
exhibits somewhat different Drude dc limit 
(6300$\pm$540\,$\Omega\cdot$cm$^{-1}$) and $\gamma_D$ (1.2$\pm$0.1\,eV), see Table\,\ref{Table1} and Figure\,\ref{Drude}, which can be attributed to the discontinuous nanoisland structure of this layer.

Finally, we would like to note that it will be challenging 
to investigate dc transport and superconductivity properties of 
the ultrathin Bi films possessing 2D TI surface states 
following the approach presented in Ref.\,\cite{Suslov}, 
where the subkelvin superconductivity without any external stimuli 
was discovered in 3D TI Cd$_3$As$_2$ films \cite{Kochura,Kovaleva_metals_2020}. 
   

\section{Conclusions}
In summary, using wide-band (0.5-6.5\,eV) spectroscopic ellipsometry we 
studied the optical properies of the 
[Bi(0.6,\,1.4,\,2.0,\,2.5\,nm)--NiFe(1.8\.nm)]$_{16}$ MLFs 
prepared by rf sputtering. 
The XRD analysis suggested that the 0.6,\,1.4,\,2.0,\,and 2.5\,nm Bi layers in the studied MLFs correspond to about two, four, six, and eight 
Bi$\left(012 \right)$-type monolayers, respectively. From the multilayer model simulations of the measured ellipsometric data, we extracted the Bi and FeNi layer dielectric functions. The dielectric function for the 2.0 and 2.5\,nm Bi spacer layers are represented by the Drude resonance due to the surface states and the low-energy Lorentz band peaking at around 0.3-0.4\,eV. The pronounced Lorentz band is characteristic of the semimetallic bulk-like Bi electronic zone structure due to the contributions from the interband transitions near the $\Gamma$ point. We discovered that the 2.0 and 2.5\,nm Bi spacer layers can be associated with the semimetallic Bi phase sandwiched between two trivial (where the topology number $\nu$=0) metallic layers on the top and bottom surfaces. The contribution from the low-photon-energy Lorentz band is strongly suppressed for the 1.4 and 0.6\,nm Bi layers, where the Drude conductivity displays notably improved metallicity properties. This indicates that the Bi layer consisting of 4 Bi$\left( 012 \right)$-type monolayers represents a kind of crossover regarding the contributions from the semimetallic Bi phase and/or surface metallic-like states. Therefore, the properties of Bi layers below 4 monolayers may be associated with nontrivial topology (where the topology number $\nu$=1) peculiar for an intrinsic 2D TI.  
We expect that the Bi layers having thickness of 0.9\,nm will exhibit maximal GMR effect of about 20\% in the (Bi-FeNi) MLFs, where the Drude dc limit is about 8970$\pm$540\,$\Omega\cdot$cm$^{-1}$. These states may be protected from backscattering, which makes them promising in spintronic devices and quantum computing. \\

{\bf Acknowledgement}

We thank F.A. Pudonin for providing us with the 
Bi/FeNi multilayer film samples and O. Pacherova for their 
XRD analysis. We thank A. Muratov for participation in the spectroscopic ellipsometry measurements. This work was supported by the European Structural and Investment Funds and the Czech Ministry of Education, Youth, and Sports (Project No. SOLID21, Cz.02.1.01/0.0/0.0/16$_{-}$019/0000760).\\  
 
{\bf Declaration of competing interest}

The authors declare no conflict of interest.


\begin{thebibliography}{99}

\bibitem{Rashba}Bychkov, Y.A.; Rashba, E.I. {\em JETP Lett.}, {\bf 1984}, {\em 39}, 78. 
 
\bibitem{Golin}Golin, S. {\em Phys. Rev. B}, {\bf 1968}, {\em 166}, 643.

\bibitem{Gonze}Gonze, X.; Michenaud, J.-P.; Vigneron, J.-P. {\em Phys. Rev. B}, {\bf 1990}, {\em 41}, 11827.

\bibitem{Liu1}Liu, Y.; Allen, R.E. {\em Phys. Rev. B}, {\bf 1995}, {\em 52}, 1566. 

\bibitem{Hofmann}Hofmann, Ph. {\em Prog. Surf. Sci.}, {\bf 2006}, {\em 81}, 191.

\bibitem{Yokota} Yokota, Y.; Takeda, J.; Dang, C.; Han, G.; McCarthy, D.N.; Nagao, T.; Hishita, S.; Kitajima, K.; Katayama, I. {\em Appl.Phys. Lett.}, {\bf 2012}, {\em 100}, 251605.

\bibitem{Hoffman}Hoffman, C.A.; Meyer, J.R.; Bartoli, F.J. {\em Phys. Rev. B}, {\bf 1993}, {\em 48}, 11431. 

\bibitem{Koroteev} Koroteev, Yu. M.; Bihlmayer, G.; Chulkov, E.V.; Blugel, S. {\em Phys. Rev. B}, {\bf 2008}, {\em 77}, 045428.  

\bibitem{Wada}Wada, M.; Murakami, S.; Freimuth, F.; Bihlmayer, G. 
{\em Phys.Rev.B}, {\bf 2011}, {\em 83}, 121310(R).

\bibitem{Murakami} Murakami, S. 
{\em Phys. Rev. Lett.}, {\bf 2006}, {\em 97}, 236805. 

\bibitem{Kane}Fu, L.; Kane, C.L.; Mele, E.J. {\em Phys. Rev. Lett.}, {\bf 2007}, {\em 98}, 106803.  

\bibitem{Liu2}Liu, Z.; Liu, C.-X.; Wu, Y.-S.; Duan, W.-H.; Liu, F.; Wu, J. 
{\em Phys. Rev. Lett.}, {\bf 2011}, {\em 107}, 136805. 

\bibitem{Kovaleva_APL_2021}Kovaleva, N.N.; Chvostova, D.; Pacherova O.; Muratov A.V.; Fekete L.; Sherstnev I.A.; Kugel K.I.; Pudonin F.A.; Dejneka A. 
{\em Appl. Phys. Lett.}, {\bf 2021}, {\em 119}, 183101. 

\bibitem{Sherstnev} Sherstnev, I.A. 
{\em Ph. D. Thesis}; P.N. Lebedev Physical Institute: Moscow, Russia, 2014.

\bibitem{Boltaev} Boltaev, A.P.; Pudonin, F.A.; Shertnev, I.A.; 
Egorov, D.A. 
{\it JETP}, {\bf 2017}, {\em 125}, 465.

\bibitem{VASE}Woollam, J.A. {\em VASE Spectroscopic Ellipsometry Data Analysis Software}; J.A. Woollam, Co.: Lincoln, NE, 2010.

\bibitem{Stupakov}Stupakov, A.; Bagdinov, A.V.; Prokhorov, V.V.; 
Bagdinova, A.N.;  Demikhov, E.I.; Dejneka, A.; Kugel, K.I.; 
Gorbatsevich, A.A.; Pudonin, F.A.; Kovaleva, N.N. 
{\em J. Nanomater.}, {\bf 2016}, Article ID 3190260. 

\bibitem{Kovaleva_APL_2015}Kovaleva, N.N.; Chvostova, D.; 
Bagdinov, A.V.; Petrova M.G.; Demikhov E.I.; Pudonin F.A.; 
Dejneka A. 
{\em Appl. Phys. Lett.}, {\bf 2015}, {\em 106}, 051907. 

\bibitem{Kovaleva_metals}Kovaleva, N.; Chvostova, D.; Dejneka, A. 
{\em Metals}, {\bf 2017}, {\em 7}, 257.  

\bibitem{Palik}Palik, E.D. {\em Handbook of Optical Constants of Solids}; Elsevier Science: USA, 1991.

\bibitem{Huetten} H\"utten, A.; Mrozek, S.; Heitmann, S.; Hempel, T.; Br\"uckl H.; Reiss, G. {\em Acta mater.}, {\bf 1999}, {\em 47}, 4245.

\bibitem{Mathon}Mathon, J. {\em Contemporary Physics}, {\bf 1991}, {\em 32}, 143.

\bibitem{Nagao} Nagao, T.; Sadowski, J.T.; Saito, M.; Yaginuma, S.; Fujikawa, Y.; Kogure, T.; Ohno, T.; Hasegawa, S.; Sakurai, T. {\em Phys. Rev. Lett.}, {\bf 2004}, {\em 93}, 105501. 

\bibitem{Wang}Wang, P.Y.; Jain, A.L. {\em Phys. Rev. B}, {\bf 1970}, {\em 2}, 2978.

\bibitem{Lenham}Lenham, A.P.; Treherne, D.M.; Metcalfe, R.J. {\em J. Opt. Soc. Am.}, {\bf 1965}, {\em 55}, 1072. 

\bibitem{Hunderi}Hunderi, O. {\em J. Phys. F}, {\bf 1975}, {\em 5}, 2214.

\bibitem{Toudert}Toudert, J.; Serna, R. {\em Opt. Mater. Express}, {\bf 2017}, {\em 7}, 2299.
 
\bibitem{Kovaleva_SciRep}Kovaleva, N.N.; Kusmartsev, F.V.; Mekhiya, A.B.; Trunkin, I.N.; Chvostova, D.; Davydov, A.B.; Oveshnikov, L.N.; Pacherova, O.; Sherstnev, I.A.; Kusmartseva, A.; Kugel, K.I.; Dejneka, A.; Pudonin, F.A.; Luo,Y.; Aronzon, B.A. 
{\em Sci. Rep. }, {\bf 2020}, {\em 10}, 21172.

\bibitem{Suslov}Suslov, A.V.; Davydov, A.B.; Oveshnikov, L.N.; Morgun, L.A.; Kugel, K.I.; Zakhvalinskii, V.S.; Pilyuk, E.A.; Kochura, A.V.; Kuzmenko, A.P.; Pudalov, V.M.; Aronzon, B.A. 
{\em Phys. Rev. B}, {\bf 2019}, {\em 99}, 094512.   

\bibitem{Kochura}Kochura, A.V.; Zakhvalinskii, V.S.; Htet, A.Z.; Ril', A.I.; Pilyuk, E.A.; Kuz'menko, A.P.; Aronzon, B.A.; Marenkin, S.F. 
{\em Inorg. Mater.}, {\bf 2019}, {\em 55}, 879.

\bibitem{Kovaleva_metals_2020}Kovaleva, N.; Chvostova, D.; Fekete, L.; 
Muratov, A. {\em Metals}, {\bf 2020}, {\em 10}, 1398.

\end{thebibliography}
\end{document}